\documentclass[aps,amssymb,twocolumn,superscriptaddress,unsortedaddress,nofootinbib]{revtex4}
\usepackage{setspace}
\usepackage{graphicx}
\usepackage{psfrag}
\usepackage[dvips]{color}

\def\be{\begin{equation}}
\def\ee{\end{equation}  }
\def\bea{\begin{eqnarray}}
\def\eea{\end{eqnarray}  }

\newcommand{\secintro}{I}
\newcommand{\secmeth}{II}
\newcommand{\secresults}{III}
\newcommand{\secconclusion}{IV}

\begin{document}
\title{Ultra Relativistic Particle Collisions}

\author{Matthew W. Choptuik}
\affiliation{CIFAR Cosmology and Gravity Program \\
     Department of Physics and Astronomy,
     University of British Columbia,
     Vancouver BC, V6T 1Z1 Canada}
\author{Frans Pretorius}
\affiliation{Department of Physics, Princeton University, Princeton, NJ 08544, USA.}

\begin{abstract}
We present results from numerical solution of the Einstein field equations
describing the head-on collision of two solitons boosted to ultra 
relativistic energies. We show, for the first time, that at sufficiently
high energies the collision leads to black hole formation, consistent
with hoop conjecture arguments. This implies that the non-linear 
gravitational interaction between the kinetic energy of the solitons
causes gravitational collapse, and that arguments for black hole
formation in super-Planck scale particle collisions are robust.
\end{abstract}

\maketitle

\noindent{\bf{\em \secintro. Introduction.}}
Using ultra relativistic scattering of particles to probe the nature
of the fundamental forces has a long tradition in modern physics. 
One reason why is the de Broglie relation, stating that the characteristic
wavelength of a particle is inversely related to its momentum,
and consequently probing short-range interactions between particles
requires large momenta. In any conventional setting gravity is
an irrelevant force in such interactions. However in general
relativity all forms of energy, including momentum, gravitate, and
thus at sufficiently high energies one would expect gravity to become
important. The current paradigm suggests that this will happen
when center of mass energies approach the Planck scale, and for collisions
with energies sufficiently above this, that black holes will be 
formed~\cite{bhprod}.

The four dimensional Planck energy $E_p$ is $\approx 10^{19}$ GeV, wholly out
of reach of terrestrial experiments, and as far as known is not reached
by any astrophysical process, barring the Big Bang or the unobservable regions
inside black holes. However, if there are more than 4 
dimensions, intriguing scenarios have been suggested where the true
Plank scale is very different from what is then just an effective
4-dimensional Planck scale~\cite{large_extras}. A ``natural''
choice for the true Planck energy is the electroweak scale of $\sim$ TeV,
as this would solve the hierarchy problem. If this were the case,
and the paradigm of black hole formation is correct, this would imply
that the Large Hadron Collider (LHC) will produce black holes, and that
black holes are formed in the earth's atmosphere by cosmic rays~\cite{BH_prod}.
Existing experimental bounds on the Planck energy in this context are at around 
1 TeV~\cite{Yao:2006px}.

However, one potential problem with the above scenario, even before one
considers issues regarding the existence of extra dimensions, 
physics near the Planck regime, etc., is whether in
classical general relativity the generic outcome of ultra relativistic two ``particle'' 
scattering is a black hole for small impact parameters. There have so far been 
no solutions to the field equations demonstrating this, and as we will outline
next, the evidence usually presented for the case of black hole
formation is based on a set of conjectures and the use of limiting-case
solutions of dubious applicability.

The main argument for black hole formation is a variant of Thorne's hoop 
conjecture~\cite{thorne}: if a total amount of matter/energy $E$ is compressed
into a spherical region such that a hoop of proper circumference $2\pi R$
completely encloses the matter in all directions, a black hole will form if
the corresponding Schwarzschild radius $R_s=2GE/c^4$ is greater than $R$,
where $G$ is Newton's constant and $c$ the speed
of light. To apply this to the collision of two classical spherical solitons,
each with rest mass $m_0$ and traveling toward each other with speed $v$ in the center
of mass frame, let $E=2\gamma m_0 c^2$, the total energy of the system,
where $\gamma=1/\sqrt{1-v^2/c^2}$. The largest radius $R$ to be
enclosed by the hoop is the rest-frame radius $R_0$, as
the Lorentz contraction only flattens the particles in the direction of
propagation. The hoop conjecture then says black holes will form if
$\gamma \gtrsim c^4R_0/4Gm_0$.

The above argument is purely classical. Quantum mechanics enters
with the {\em assumption} that the argument still holds for the
collision of fundamental particles, now taking $R_0$ to be
the DE Broglie wavelength $hc/E$ of the particle, where $h$
is Planck's constant. Dropping constants of $O(1)$, the criteria for 
black hole formation is then $E \gtrsim (hc^5/G)^{1/2}$, the Planck energy.
However, our goal in this paper is only to address the 
soundness of the {\em classical} hoop-conjecture argument; if it fails,
there is no reason to expect a full quantum version to hold.

It is not obvious that the hoop conjecture is applicable in all situations.
Consider a single particle boosted beyond the Planck energy. Since the boosted particle's 
spacetime is a coordinate transformation of its rest frame geometry, there is clearly
no black hole formation. As trivial as this example may seem, it is still insightful,
as it illustrates that not all forms of energy gravitate in the same way.
Here, kinetic energy, unlike rest mass energy, does not 
produce spacetime curvature, yet both forms of energy contribute 
to the mass of the spacetime, as measured for instance by the ADM mass~\cite{ADM}.
The kinetic energy dominates the rest mass energy by orders of magnitude, and
for black hole formation to be a {\em generic} outcome
the particular nature of the particles and non-gravitational interactions between them
cannot play a role.
Therefore, it must be the non-linear {\em interaction} between opposing
streams of gravitational kinetic energy that causes a black hole to form, in
the process converting kinetic to rest mass and gravitational wave
energy. 

What is oft-quoted as evidence for black hole formation comes from the
study of the collision of two infinitely boosted ``particles'', described
by the Aichelburg-Sexl (AS) metric~\cite{Aichelburg:1970dh}. The AS
solution is obtained by taking a Schwarzschild black
hole of mass $m$, applying a Lorentz boost $\gamma$, and then taking
the limits $\gamma\rightarrow\infty$ and $m\rightarrow0$, so that 
the product $E=\gamma m$ remains finite. The result is a gravitational
``shock wave'', where the non-trivial
geometry is confined to a 2-dimensional plane traveling at the speed
of light, with Minkowski spacetime on either side.
Two such solutions, moving in opposite directions,
can be superimposed to give the pre-collision geometry of the spacetime (see~\cite{D'Eath:1976ri}
for an insightful description). Though the geometry
is not known to the future of the collision, at the moment
of collision a trapped surface can be found~\cite{urbh}.  Assuming 
cosmic censorship, this would be an example of black
hole formation in an ultra relativistic collision.

There are several aspects of the infinite boost construction that
should give one pause as to its applicability to a
large-yet-finite $\gamma$ collision of massive particles.
The AS limit is not asymptotically flat, and the algebraic type of the
metric has changed from Petrov type D (two distinct null eigenvectors
of the Weyl tensor) to Petrov type N (one null eigenvector). This latter
point can be thought of as the gravitational field changing from
a Coulomb-like to a pure gravitational wave field.
The AS metric is also not a good description to the geometry of
a finite boosted particle on the shock surface;
one is then left with
the un-insightful conclusion that the description is good
sufficiently far from the particle that its metric is Minkowski. 
It has also been argued that black hole formation is due to the strong focusing 
of geodesics off the AS shockwave~\cite{Kaloper:2007pb}. However, there is no dynamics in this
description (neither in the superposed AS metrics for that matter), and
it is difficult to imagine how the geodesic
structure can capture what is a highly non-linear
and dynamical interaction between gravitational energy.

To test the hypothesis that black holes form in high energy collisions
of particles, we numerically solve the Einstein field equations coupled to matter
that permits stable, self gravitating soliton solutions---these are our model particles.
The particular soliton we use is a {\em boson star}~\cite{bsrefs}.
One motivation for choosing this model was from earlier studies of low velocity, head-on
collisions of boson stars, which suggested that as the velocity
increases, gravity appears to ``weaken'', and the boson stars pass through each other
exhibiting a non-relativistic solitonic interference pattern~\cite{choi_nbs}. In particular,
the magnitude of the bosonic matter field developed interference fringes
of wavelength $\lambda\propto 1/P$, with $P$ the momentum of each boson star, which is
exactly the relationship observed during the collision of Bose-Einstein condensates
bound via {\em Newtonian} gravity.
This model therefore seems perfect to address the genericity requirement for 
black hole formation at super-Planck
scale collisions, in that the self interaction of the matter will not
bias the outcome toward black hole formation 
(as, for example, using black holes as model particles would).

\noindent{\bf{\em \secmeth. Methodology.}}
We solve the Einstein equations, 
$R_{ab}-g_{ab}/2 R=8\pi T_{ab}$,
using a variant of the generalized harmonic formalism~\cite{ghform} with
constraint damping~\cite{gundlach_et_al} as described in~\cite{paper12}.
In this formalism, the spacetime coordinates satisfy a wave (harmonic) condition $\Box x^a = H^a$, with
the $H^a$ encoding the coordinate (gauge) degrees of freedom.
Experimentation with coordinate choices led us to develop what we term 
a {\em damped harmonic} condition, which can be written 
$H^a = \xi \left[n^a-\bar{n}^a\right]$ where $\xi$ is a constant,
$n^a$ is the time-like unit normal to the $t=\rm const.$ slices, and
$\bar{n}^a$ is another time-like unit normal field which is to be chosen
so that the resulting coordinate system is non-singular.  Such an approach
was independently introduced in~\cite{Lindblom:2009tu}, where the 
choice $\bar{n}^a = (\partial/\partial t)^a/\alpha + \log(\alpha/\sqrt{h}) n^a$,
with $\alpha$ the lapse function and $h$ the determinant of the spatial
metric, was proposed.  We use a variant of this condition that transitions
to $H^a=0$ shortly after the collision.

The matter is a minimally coupled complex scalar field $\phi$
with mass parameter $m$,
equation of motion $\Box\phi=m^2 \phi$, and 
stress tensor 
$T_{ab} = 2 \nabla_{[a}\phi \nabla_{b]} \bar{\phi} -
g_{ab} \left( \nabla_c \phi \nabla^c \bar{\phi} + m^2\phi\bar{\phi} \right)$,
where $\bar{\phi}$ is the complex conjugate of $\phi$. 
For initial data we superimpose two boosted boson stars
following a procedure analogous to that for binary black holes
presented in~\cite{Matzner:1998pt}.
This construction by itself does not fully satisfy the
constraint equations. However, 
the further apart the boson stars are at $t=0$, the smaller the error is,
and we have performed tests that indicate for the initial separations here
the error is sufficiently small to not qualitatively affect the conclusions.

\begin{figure*}
\begin{tabular}{@{\extracolsep{1.0in}}cccc}
 $\gamma=1$ & $\gamma=1.15$ & $\gamma=2.75$ & $\gamma=4$
\end{tabular}
\vspace{-0.125in}
\begin{center}
\includegraphics[width=3.85cm,clip=false]{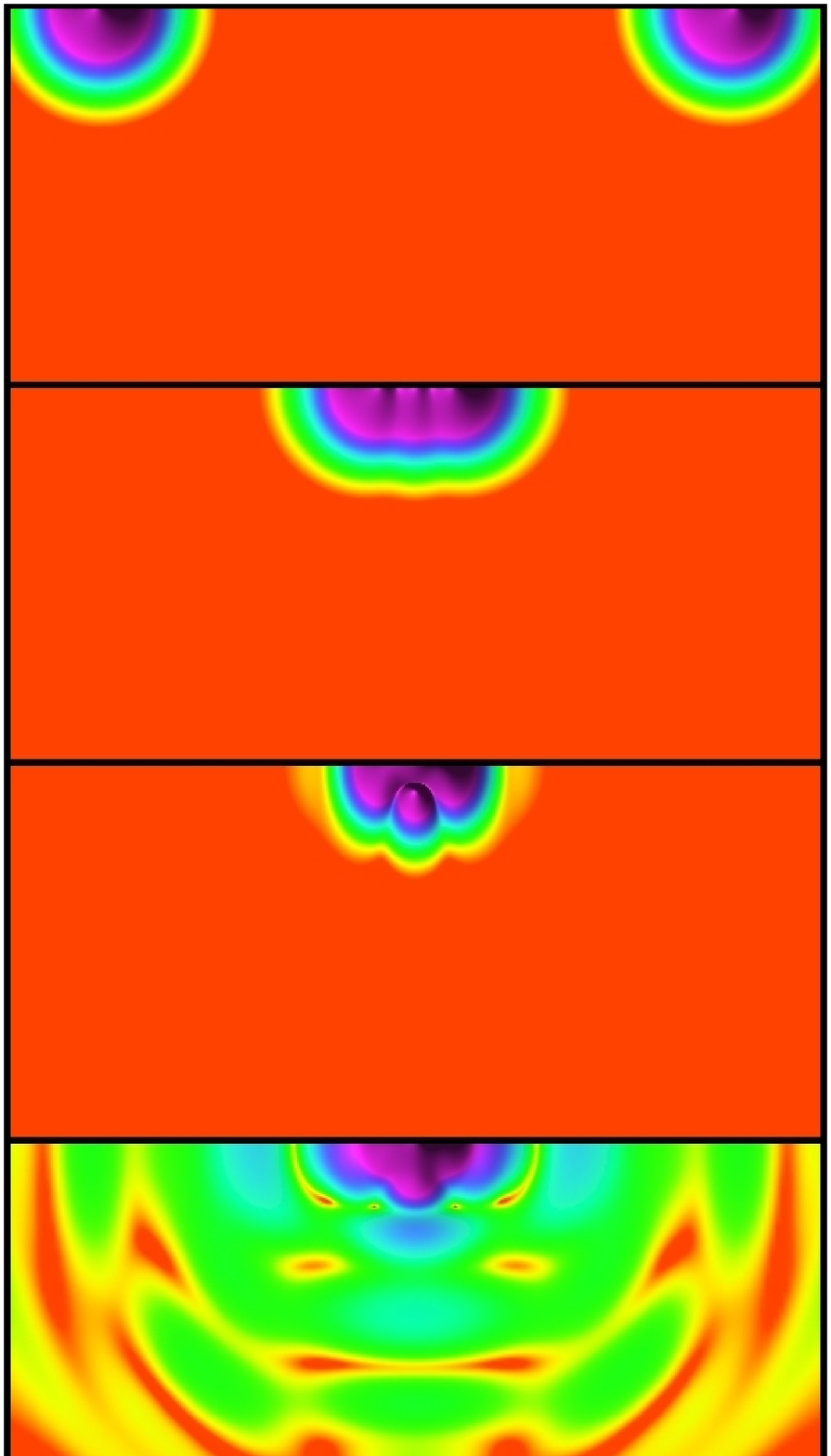}
\includegraphics[width=3.85cm,clip=false]{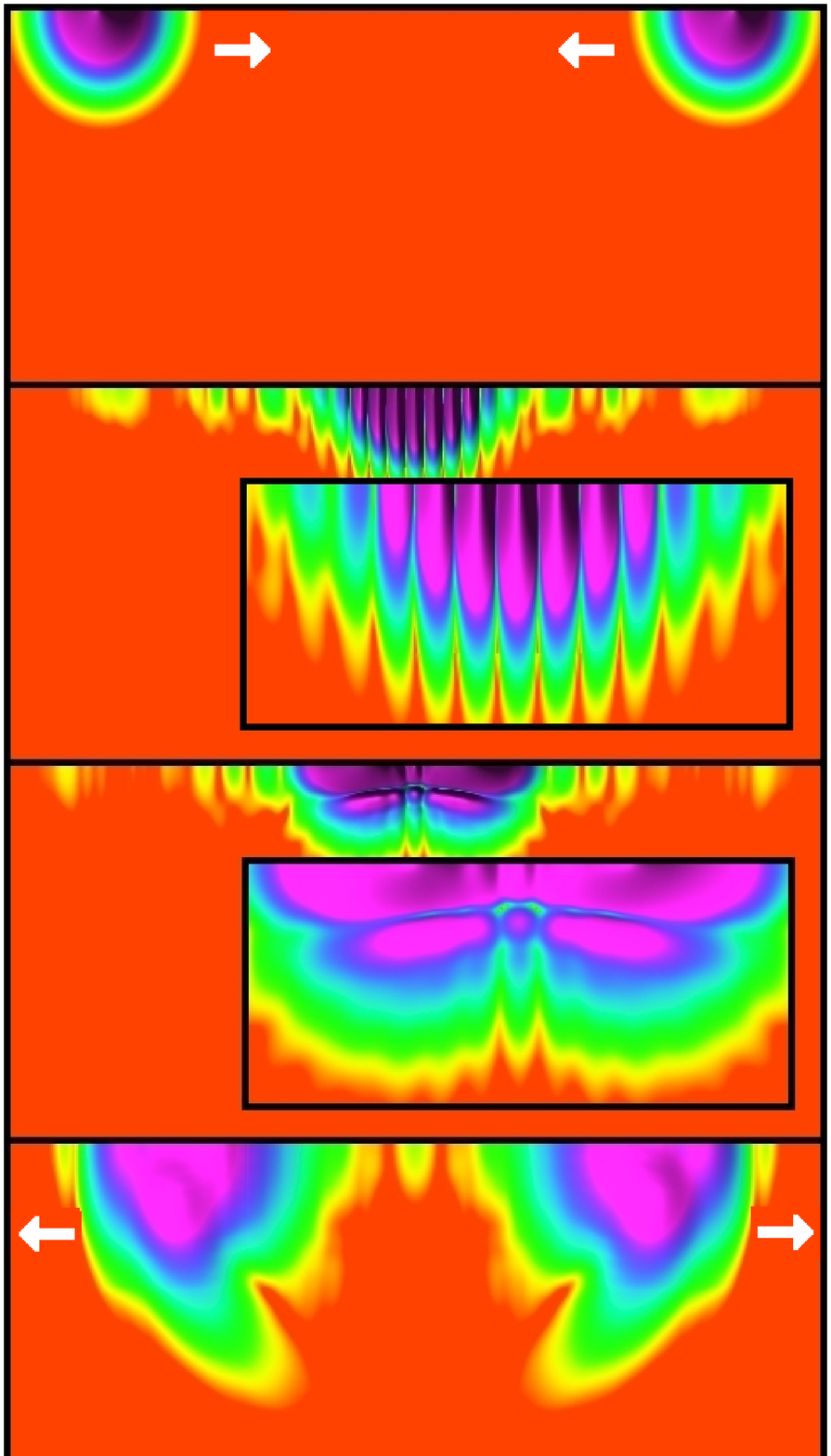}
\includegraphics[width=3.85cm,clip=false]{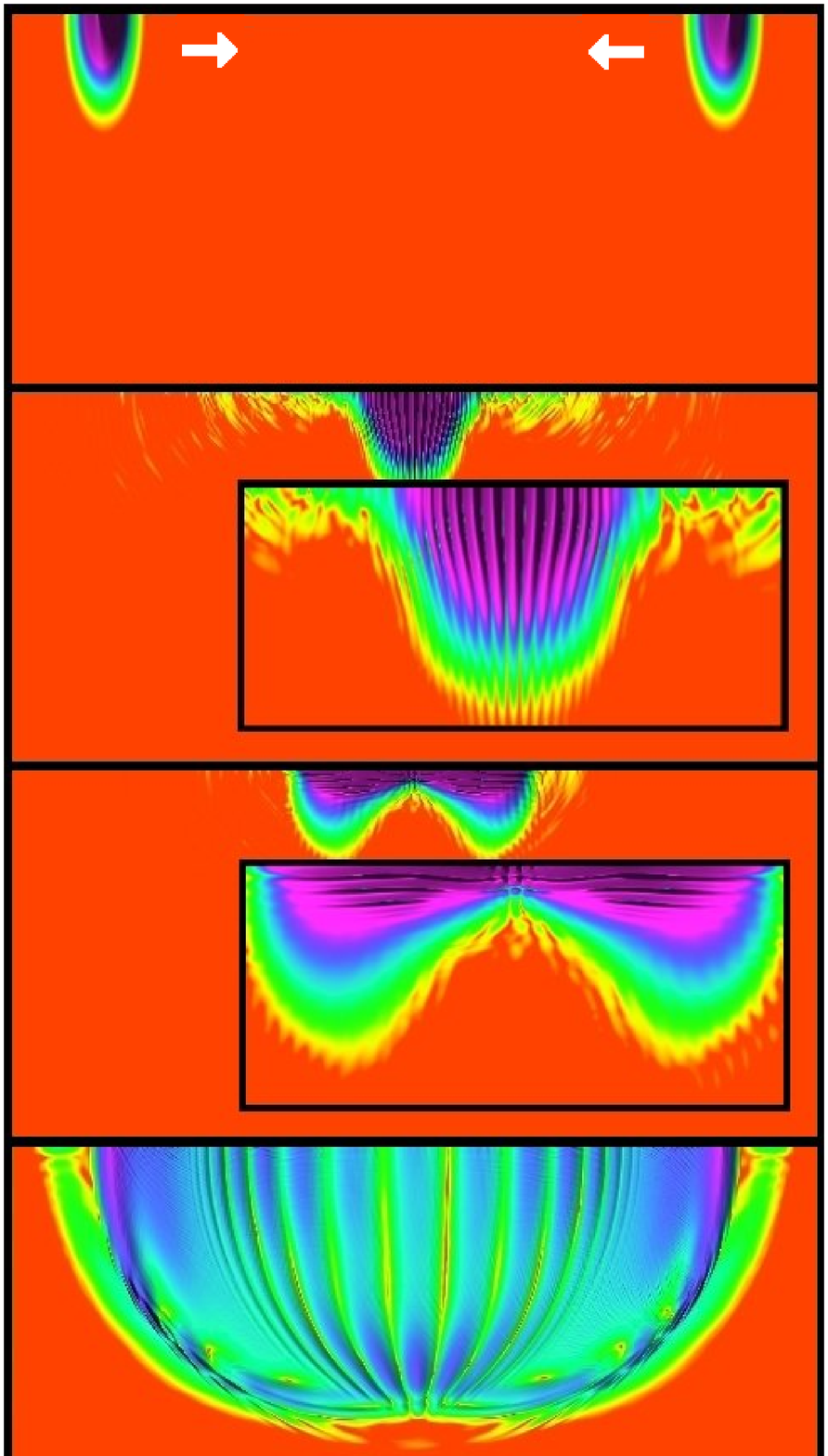}
\includegraphics[width=3.85cm,clip=false]{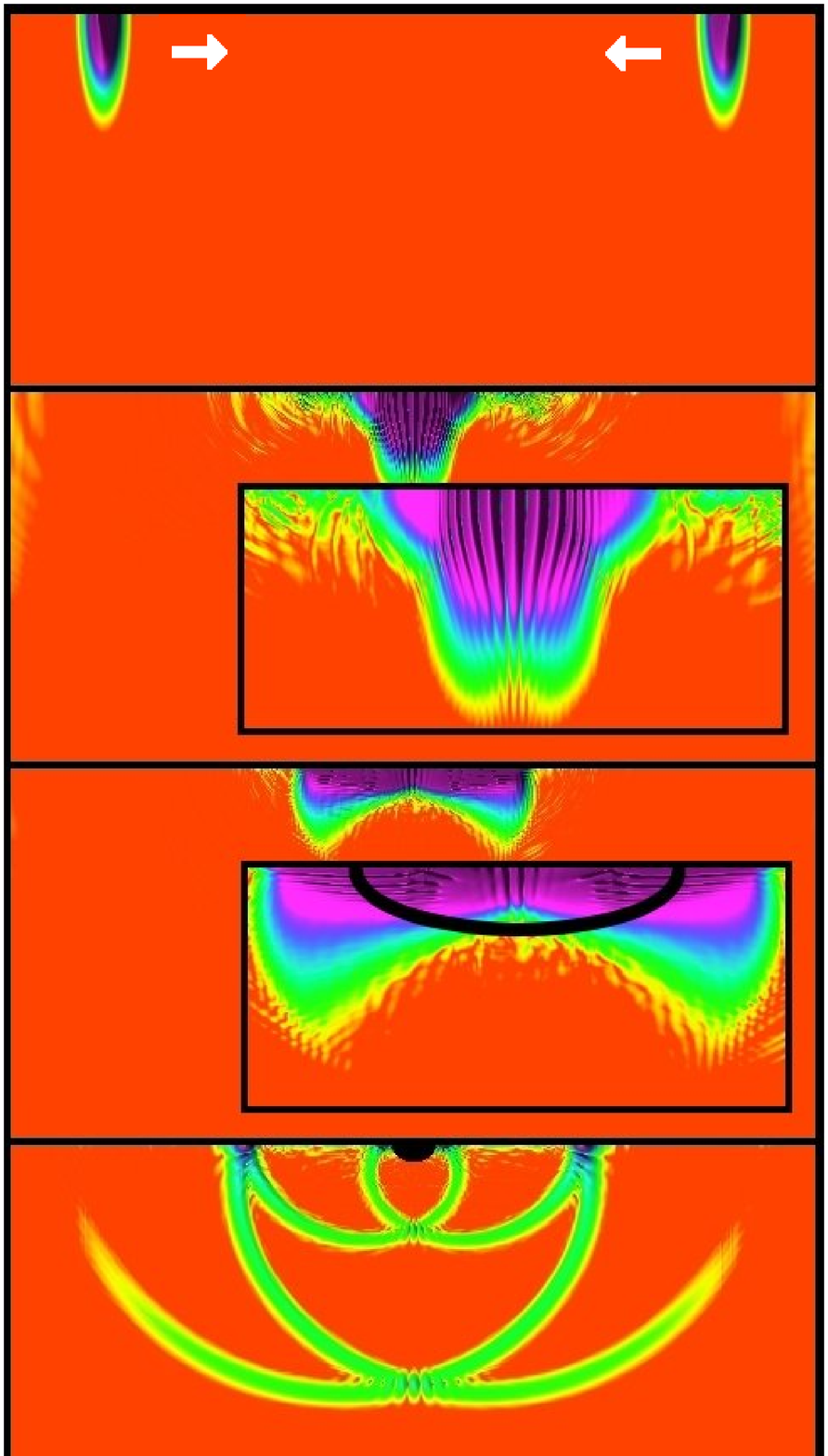}
\end{center}
\caption{Magnitude $|\phi|$ of the scalar field from 4 different
simulations, in 4 panels (left to right). The 4 sub-panels within each 
panel depict $|\phi|$ at different times as follows (top to bottom): 
1) $t/M_0=0$, 2) a time at which the boson stars first completely overlap,
3) a short time later when $|\phi|$ reaches a first local maximum due 
to gravitational focusing, 4) a late time after the collision. 
The axis of symmetry is coincident with the top edge of each sub-panel.
The insets, where present, are zoom-ins of the 
central interaction regions.
For the $\gamma=4$ case, a black hole
forms near the time of sub-panel 3---the black line in the corresponding inset shows
the shape of the apparent horizon then, and
the black semi-circle in sub-panel 4 is the excised region inside the black hole.
}
\label{fig_gammas}
\end{figure*}

Each boson star is identical with a central scalar field amplitude $\phi_0$
chose so that the maximum compactness $2 M(r)/r$ of
each star is $\approx1/20$, where the mass aspect, $M(r)$,
approaches the ADM mass $M_{\rm ADM}$ as $r\rightarrow\infty$.
We subsequently scale all units to $M_0 \equiv 2 M_{\rm ADM}$.
We choose an initial coordinate separation between the boson
stars in the center of mass (simulation) frame 
of $d_0=250 M_0$, and give each boson star a boost of $\gamma$.
Thus, with this compactness, the hoop-conjecture estimate of the
black hole formation threshold is $\gamma_h\approx 10$.

\noindent{\bf{\em \secresults. Results.}}
Our key result is the simple answer {\em ``yes!''} to the question of 
{\em ``do ultra relativistic boson star collisions lead to black hole 
formation in classical general relativity?''}. Despite the objections
we noted to the use of the hoop conjecture in
this scenario, it therefore does appear that the argument captures
the essential physics of high speed soliton collisions. 
We find black hole formation at $\gamma_c=2.9\pm10\%$, {\em less} than a third that 
predicted by the hoop conjecture. Of course, the latter
is an order of magnitude estimate, and the numerical value of the threshold may
depend on the particular soliton model. Note that 
the maximum compactness of single, stable boson stars is 
$\sim 0.25$ (see for e.g.~\cite{Hawley:2000dt}), though it is not immediately
apparent whether this is of relevance here.


Fig.~\ref{fig_gammas} shows snapshots of the 
scalar field for several boost parameters 
at key times. For the collision beginning at rest ($\gamma=1$),
a single perturbed boson star forms, undergoing large oscillations
that slowly damp via the emission of scalar radiation.
For larger boosts, the initial boson star interaction exhibits the 
usual non-gravitational interference pattern, but shortly afterward 
there is some compression of the stars due to gravity. For the modest boost 
of $\gamma=1.15$, though the stars are perturbed by the compression, they pass 
through each other. Approaching the threshold with $\gamma=2.75$, the compression 
is much greater, and though the boson stars pass through each other the 
perturbation is strong enough to cause them to ``explode''. 
I.e., though the bulk of the momentum in the scalar field is
concentrated in two fronts propagating outward along the axis, 
a non-negligible component appears to move outward in spherical shells emanating 
from {\em two} focal points, corresponding to the locations of maximum 
compression seen in sub-panel 3.

For the $\gamma=4$ case the interaction is similar to $\gamma=2.75$ until 
apparent horizon formation; this is consistent with the intuition that in this 
regime the ``matter does not matter''---it is the gravitational energy
determining the dynamics, and here the scalar field is merely a tracer
of the underlying geometry. After black hole formation, the resultant
evolution is strikingly different. Most of the scalar field
falls into the black hole, though a small fraction escapes. Similar
to the $\gamma=2.75$ case, at late times (sub-panel 4) the matter that escapes 
can be traced back to appear to have originated as two pulses on the axis at
the locations of maximum compression (sub-panel 3). However, now a piece
of each outward moving wavefront gets trapped in the black hole. The wavefront
remains connected, and with time this causes the concentric outward 
propagating arc patterns seen in sub-panel 4.

\begin{figure}[t]
\begin{center}
\includegraphics[width=8.0cm,clip=false]{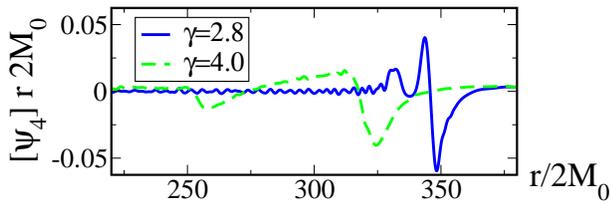}
\end{center}
\caption{
$\Psi_4$ at $t=540M_0$ on the plane
passing through the collision point and orthogonal to the axis,
for two cases.
}
\label{fig_gw}
\end{figure}

Fig.~\ref{fig_gw} depicts the gravitational
wave emission, as measured by the Newman-Penrose scalar $\Psi_4$,
for the two higher $\gamma$ cases of Fig.~\ref{fig_gammas}.
Note however that the damped harmonic coordinates cause a strong
``distortion'' of the metric near the solitons, which eventually
propagates outward with the gravitational wave. This 
prevents a clean interpretation of $\Psi_4$, defined here with
a tetrad aligned with the coordinate basis vectors,
as representing {\em the} gravitational wave signal.
With this caveat in mind, there is an interesting feature suggested
by Fig.~\ref{fig_gw}. The wavelength in the black hole formation
case is consistent with the wave being associated with the 
dominant quasi-normal ringdown mode of the black hole. However,
in the sub-threshold case the characteristic wavelength
is quite a bit shorter, and there seems to be a trend of smaller
length-scale features developing the closer to threshold we tune,
which will continue if the threshold exhibits type II critical
behavior~\cite{Choptuik:1992jv}.
As it relates to particle collisions, this suggests black hole formation
may not be the ``end of short-distance physics'', but that the low energy
relationship equating small distances to large momenta ceases to be
valid, and probing physics at smaller distances
requires fine-tuning the interaction energy.

\noindent{\bf{\em \secconclusion. Conclusions.}}
We presented numerical results from a first study of the ultra relativistic 
collision of solitons within general relativity.
The goal was to test if at sufficiently
high energy gravity dominates the interaction, leading to black hole
formation. We found that, for this class of soliton, the conjecture
is true, and the threshold of black  hole formation occurs
at a boost $\gamma_c$ approximately $1/3$ that predicted
by Thorne's hoop conjecture. Interestingly, a factor of 
$\sim 1/3$ also arises in calculations of trapped-surface formation
in the collision of null sources following an S-matrix approach
to the scattering problem~\cite{transplanck}.
With $\gamma_c = 2.9 \pm 10\%$
the ratio of kinetic to rest mass energy is $\approx 2:1$; we believe
this is sufficiently large to make a compelling case that black hole
formation is generic in ultra relativistic particle collisions, regardless
of the internal structure of the particles. Thus the arguments that
super-Planck scale particle 
collisions lead to black hole formation
are robust, and furthermore using black holes as the model
particle to study the gravitational aspects of the interaction
at these energies~\cite{D'Eath:1976ri,urbh,urbhs} is valid.
Our results also suggest that when gravity becomes
a strong player in the interaction at scales slightly below the Planck
scale, the de Broglie relationship equating smaller distances to larger
energies may cease to hold. 
Of course, here the nature of quantum gravity will be crucial,
and may provide its own cut-off to short distance physics, though 
gravity will not play the role of the censor.

\noindent{\bf{\em Acknowledgments.}}
We thank Lee Lindblom, Bela Szilagyi, Don Page and Gabriele Veneziano for 
useful comments. 
This work was supported by NSERC (MWC), CIFAR (MWC), 
the Alfred P. Sloan Foundation (FP), and NSF grant PHY-0745779 (FP).
Simulations were run on {\bf Woodhen} (Princeton University) and
{\bf Glacier} clusters (Westgrid, supported by CFI, ASRI and BCKDF).

\end{document}